\documentclass[conference]{IEEEtran}

\usepackage{amsmath}
\usepackage{amsfonts}
\usepackage{geometry}
\usepackage{amsthm}
\usepackage{graphicx}
\usepackage{mathtools,color}
 
\makeatletter
\numberwithin{figure}{section}
\theoremstyle{plain}
\newtheorem{thm}{\protect\theoremname}
  \theoremstyle{plain}
  \newtheorem{lem}[thm]{\protect\lemmaname}
  \theoremstyle{definition}
  
  \theoremstyle{plain}
  
  \theoremstyle{plain}

  \theoremstyle{plain}
  \newtheorem{cor}[thm]{\protect\corollaryname}
\newtheorem{rem}{\protect\remarkname}
  \theoremstyle{plain}

  \theoremstyle{plain}

\@ifundefined{showcaptionsetup}{}{%
 \PassOptionsToPackage{caption=false}{subfig}}
\usepackage{subfig}
\makeatother

  \providecommand{\propname}{Proposition}
  \providecommand{\corollaryname}{Corollary}
  \providecommand{\definitionname}{Definition}
  \providecommand{\factname}{Fact}
  \providecommand{\lemmaname}{Lemma}
\providecommand{\theoremname}{Theorem}
\providecommand{\remarkname}{Remark}

\begin{document}

\title{Quantum Enhanced Classical Sensor Networks}

\author{\IEEEauthorblockN{David E. Simmons,  Justin P. Coon}
\IEEEauthorblockA{Department of Engineering Science \\University of Oxford \\Oxford, UK, OX13PJ\\
Email: \{david.simmons, justin.coon\}@eng.ox.ac.uk} 
\and
\IEEEauthorblockN{Animesh Datta}
\IEEEauthorblockA{Department of Physics \\University of Warwick \\UK\\
Email: animesh.datta@warwick.ac.uk}}

\maketitle

\begin{abstract}
The quantum enhanced classical sensor network consists of $K$ clusters of $N_e$ entangled quantum states that have been trialled $r$ times, each feeding into a classical estimation process. Previous literature has shown that each cluster can {ideally} achieve an estimation variance of $1/N_e^2r$ for sufficient $r$. We begin by deriving the optimal values for the minimum mean squared error of this quantum enhanced classical system. We then show that if noise is \emph{absent} in the classical estimation process, the mean estimation error will decay like $\Omega(1/KN_e^2r)$. However, when noise is \emph{present} we find that the mean estimation error will decay like $\Omega(1/K)$, so that \emph{all} the sensing gains obtained from the individual quantum clusters will be lost.  
\end{abstract}

 
\section{Introduction\label{sec:INTRO1}}

\subsection{Background}

	{Quantum-enhanced sensing is being increasingly envisioned for deployment in real world situations. The real world is typified by the presence of noise, in which case it is known that the performance of quantum sensors cannot provide any improved scaling~\cite{Demkowicz-Dobrzanski2012,Jarzyna2015} asymptotically. This scaling is in the number of constituent probes in the quantum sensor. For finite sensor sizes, however, quantum sensors can still outperform classical ones depending on the nature and magnitude of noise. The noise in these studies is such that it affects the quantum evolution of the sensor. }
	
	{In this work, we study an alternative scenario where a collection of quantum sensors is connected by a classical network. We take the quantum sensors to be hypothetically ideal, but consider the classical channels to be noisy. To the best of our knowledge, our is the first study of quantum sensors embedded in a classical network.
We highlight the following three key findings of this paper:}
%\subsection{Key observations}
\begin{enumerate}
\item We show that the quantum enhanced classical sensor network with $N$ elements in each entangled cluster, and $K$ clusters in total, has a best-case minimum mean squared error of $\Omega(1/KN_e^2r)$.
\item We show that when noise is present in the classical estimation portion of the quantum enhanced classical estimator, the  minimum mean squared error is given by $\Omega(1/K)$.
\item We show that the optimal performance of the network is not uniquely determined by the rank of the channel. In particular, both full and unit rank channel matrices achieve asymptotically optimal performance.
\end{enumerate}

\subsection{Document Layout}

Section \ref{sec:intro} presents background information on the benefits offered by quantum sensing approaches. The quantum enhanced classical sensor network is presented in section \ref{sec:sysmedel}, studied in section \ref{sec:QCstudy}, and discussed in section \ref{sec:study}. The paper is concluded in section \ref{sec:conc}.

\subsection{Notation}

In this work, we use $\mathbb{I}$ to denote the identity matrix, and $\mathbf{0}$ and $\mathbf{1}$ to denote vectors of $0$s and $1$s, whose sizes should be clear from the surrounding text. We use $\mathcal{I}(x)$ to denote the indicator function
\[
\mathcal{I}\left(x\right)=\left\{ \begin{array}{cc}
0 & x\leq0\\
1 & x>0.
\end{array}\right.
\]
The expectation and variance operators are denoted by $\mathbb{E}[\cdot]$ and  $\mathbb{V}[\cdot]$, respectively. We use $f(x) = \Omega(g(x))$ to mean 
\begin{equation}
\exists \;x',k \in \mathbb{R}\;\mathrm{such\;that}\;f(x) \geq |k| g(x) \;\forall\;x\geq x'.
\end{equation}
With $\{A_i\}$ being a set of equidimensional linear operators, we use $\oplus$ to denote the Kronecker sum - which is denifed by $A_1\oplus A_2 := A_1\otimes\mathbb{I}+ \mathbb{I} \otimes A_2$ - and $ \bigoplus_{k=1}^N$ to denote an iterated Kronecker sum - which is defined by 
$$ \bigoplus_{k=1}^N A_i \!:=\! A_1 \otimes \mathbb{I}\otimes\cdots\otimes \mathbb{I} + \mathbb{I}\otimes A_2 \otimes \mathbb{I}\otimes\cdots \otimes\mathbb{I} +   \mathbb{I}\otimes\cdots\otimes \mathbb{I} \otimes A_N,$$
where each summand contains $N$ tensor product factors.

\section{Background and quantum enhanced classical System Model\label{sec:intro}}

We begin by providing background information to the improvements that can be made by employing entangled quantum bits for parameter estimation, \cite{kok2004quantum,Giovannetti2011}.
Suppose we embed a phase $\phi$ onto the quantum state
\begin{equation}
\left| \chi \right> = \frac{1}{\sqrt{2}}\left( \left| 0 \right> +  \left| 1 \right>  \right)
\end{equation}
through the action of the unitary operator 
\begin{equation}
 {U} := \left| 0 \right>\left< 0\right| +  e^{\mathbf i\phi}\left| 1 \right>\left< 1 \right|.\label{eq:U}
\end{equation}
We then have 
\begin{equation}
\left| \psi \right> := {U}\left| \chi \right> = \frac{1}{\sqrt{2}}\left( \left| 0 \right> + e^{\mathbf{i}\phi}\left| 1 \right>  \right) .\label{eq:basicStaete}
\end{equation}
Suppose we then wish to estimate $\phi$ from $\left| \psi \right>$.
To measure this phase, we employ the Pauli-$X$ operator
\begin{equation}
X = \left| 0 \right>\!\left< 1 \right| + \left| 1 \right>\!\left< 0 \right| .\label{eq:SMALLSIG}
\end{equation}
The expectation of $X$ (with respect to $\left| \psi \right>$) is given by (see \eqref{eq:basicExpectation})
\begin{equation}
\left<  X \right> := \left< \psi \right| X \left| \psi \right> = \cos \phi.
\end{equation}
Repeating this experiment $N$ times yields 
\begin{equation}
\left< X_N\right> : =  \underset{N\;\mathrm{ terms }}{\underbrace{\left< \psi \right| \cdots  {\left< \psi \right|}}}\bigoplus_{k=1}^N X \underset{N\;\mathrm{ terms }}{\underbrace{ \left| \psi \right>  \cdots \left| \psi \right> }} = N\cos \phi,
\end{equation}
where $X_N$ is defined to be 
\begin{equation}
X_N= \bigoplus_{k=1}^N X.
\end{equation}
From  \eqref{eq:basicExpectation} and  \eqref{eq:sigVar}, the variance of $X$ is given by
\begin{align}
\left(\Delta X\right)^2  & :=  \left< X^2  \right> - \left< X \right>^2 = 1 - \cos^2 \phi  =  \sin^2 \phi .
\end{align}
Thus, given $N$ copies of $\left| \psi \right> $, we have 
\begin{align}
\left(\Delta X_N\right)^2 := \left< X_N^2 \right> - \left<  X_N  \right>^2 =  N \sin^2 \phi .
\end{align}
According to estimation theory \cite{sengijpta1995fundamentals}, the variance of $\hat\phi$ is given by
\begin{align}
\left( \Delta \hat\phi \right)^2 & =  \left(\frac{\Delta X_N}{\left| d\left< X_N\right> / d\phi  \right|} \right)^2= \frac{1}{N} .\label{eq:'var1/N}
\end{align}
Thus, we find that the uncertainty in the phase $\phi$ is given by the inverse of the number of samples $N$.

Suppose instead that we have an entangled system that we wish to estimate a phase $\phi$ from.  Writing $$\left|\mathbf{0}_{N_e}\right> = \underset{N_e\;\mathrm{terms}}{\underbrace{\left| 0\right> \cdots \left|0\right>}}\quad\mathrm{and}\quad\left|\mathbf{1}_{N_e}\right> =  \underset{N_e\;\mathrm{terms}}{\underbrace{\left|1\right>\cdots\left|1\right>}},$$ we let
\begin{align}
\left| \psi_e \right>  = \frac{1}{\sqrt{2}} \left(  \left|\mathbf{0}_{N_e}\right>+ e^{\mathbf{i}N_e\phi} \left|\mathbf{1}_{N_e}\right> \right).
\end{align}
As with before the phase $\phi$ has been embedded onto each of the $N_e$ quantum states through the action of $U$, \eqref{eq:U}. Our goal is to determine the measured phase $\phi$.
To do this, we need to measure the observable 
\begin{equation}
\mathcal{X} = \left|\mathbf{0}_{N_e}\right>\!\left<\mathbf{1}_{N_e}\right| +  \left|\mathbf{1}_{N_e}\right>\!\left<\mathbf{0}_{N_e}\right|.\label{eq:BIGSIG}
\end{equation}
 From \eqref{eq:basicExpectation2}, the expectation of $\mathcal{X} $ (with respect to $\left|\psi_e\right>$) is given by
\begin{equation}
\left< \mathcal{X} \right> = \cos ( N_e\phi)  .\label{eq:expectationentangledstate}
\end{equation}
Of course, because the output of the measurement with respect to $\mathcal{X}$ is binary, to measure $\phi$ (i.e., estimate \eqref{eq:expectationentangledstate}) we must repeat the measurement $r = N/N_e$ times\footnote{This choice of $N_e$ ensures that we obtain a fair comparison between the unentangled system (considiered above) and the entangled system. This is because we are considering an equal number of quantum states for both.}. 
In this scenario and for sufficiently large $r$, the variance of the phase $\phi$ is given by 
\begin{align}
\left( \Delta \hat\phi \right)^2 & =  \left( \frac{\Delta \mathcal{X}_{r}}{\left| d\left< \mathcal{X}_{r} \right> / d( \phi)  \right|} \right)^2 = \frac{ 1 }{N_e^2r} = \frac{r}{N^2} ,\label{eq:variance_entangled}
\end{align}
where 
\begin{equation}
\mathcal{X}_r = \bigoplus_{k=1}^r \mathcal{X} .
\end{equation}
Thus, by exploiting quantum entanglement we find that the uncertainty in the phase $\phi$ is decreased by a factor of $1/N_e$ relative to the unentangled system.

\subsection{System model: The quantum enhanced classical sensor network  \label{sec:sysmedel}}

\begin{figure}
{\centering{}\includegraphics[scale=0.325]{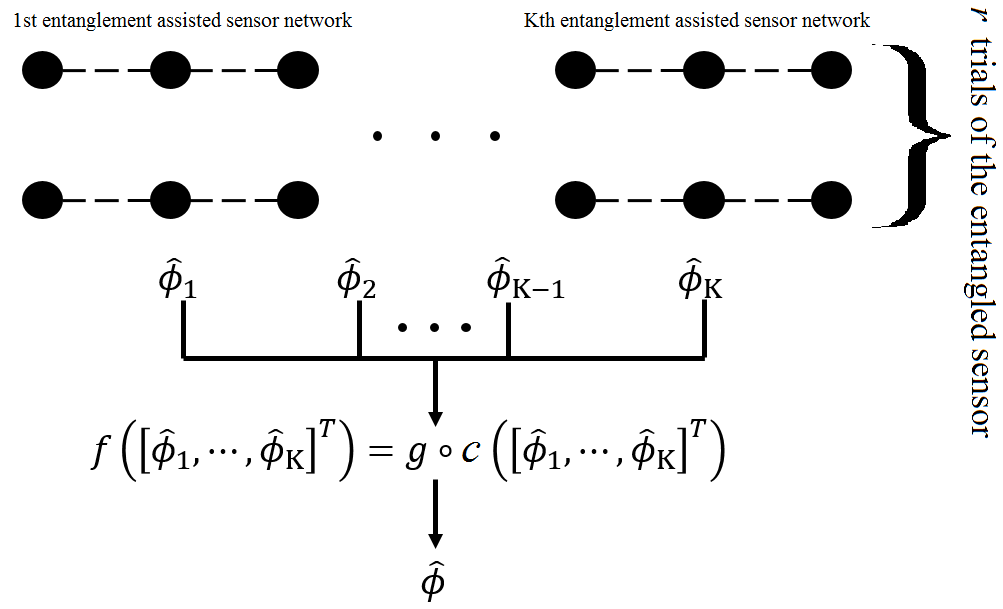}}\caption{A quantum enhanced classical sensor network. Each cluster consists of $N_e = 3$ entangled qubits, which have each been trialled $r$ times. \label{fig:syst}}
\end{figure}

Consider the quantum enhanced classical sensor network system model depicted in Fig. \ref{fig:syst}. It consists of $K$ clusters of $N_e$ entangled states (in the figure, $N_e=3$).  There exists no entanglement across distinct clusters. The $i$th cluster is described by the quantum state
\begin{align}
\left| \psi_e \right> = \frac{1}{\sqrt{2}} \left(  \left|\mathbf{0}_{N_e}\right>+ e^{\mathbf{i}N_e\phi } \left|\mathbf{1}_{N_e}\right> \right).
\end{align}
The $i$th cluster has inherited a measurement $\phi$ in the coefficient of $\left|\mathbf{1}_{N_e}\right>$.  An estimate $\hat\phi_i$ is obtained from the $i$th cluster by measuring the quantum clusters $r$ times. The estimates are \emph{independent}, \emph{identically} distributed and \emph{unbiased}. We denote the vector of estimates from all $K$ clusters by 
\begin{align}
\boldsymbol{\hat\phi} = \left[\begin{array}{cccc}\hat\phi_1&\cdots & \hat\phi_K\end{array}\right]^T.\label{eq:Kestimates}
\end{align}
From \eqref{eq:variance_entangled}, the variance of the $i$th estimate is given by
\begin{align}
\left( \Delta \hat\phi \right)^2 & = \frac{1}{{r} N_{e}^2} =  \frac{r}{N^2}.\label{eq:entangledvariance123}
\end{align}

The $K$ estimates are passed through a \emph{classical} noisy channel $c$ and combined using an estimation functional $g$ to form the final estimate $\hat\phi$ of $\phi$. The concatenation of these events (classical channel and estimation) is captured by the functional 
\begin{align}
&f=g\circ c:\mathbb{R}^K \longrightarrow \mathbb{R}\nonumber\\ \mathrm{where}\quad & g:\mathbb{R}^K \longrightarrow \mathbb{R}\quad\mathrm{and}\quad c:\mathbb{R}^K \longrightarrow \mathbb{R}^K.
\end{align}
We consider the case in which $c$ is a linear additive noisy channel, i.e., 
$$c\left(\boldsymbol{\hat\phi}  \right) = \mathbf{H}  \boldsymbol{\hat\phi} + \mathbf{n} ,$$where 
\begin{align}
\mathbf{H} & =  \left[\begin{array}{cccc}
h_{11} & h_{12} & \cdots & h_{1K}\\
h_{21} & h_{22} & \ddots & h_{2K}\\
\vdots & \ddots & \ddots & \vdots\\
h_{K1} & h_{K2} & \cdots & h_{KK}
\end{array}\right],\quad 
\mathbf{n} & = \left[\begin{array}{cccc}
n_1 \\
n_2 \\
\vdots \\
n_K
\end{array}\right] ,
\end{align}
the coefficients $h_{ij}$ are potentially non-deterministic and $n_i$ are independent zero-mean Gaussian noise terms with variance $v$. 
The functional $g$ is assumed to take the form
 $$g\left(\mathbf{H} \boldsymbol{\hat\phi}  + \mathbf{n} \right) = \mathbf{g}^T\left( \mathbf{H} \boldsymbol{\hat\phi}  + \mathbf{n}\right),$$
where $ \mathbf{g} $ is given by
\begin{align}
\mathbf{g}& = \left[\begin{array}{cccc}
g_{1} & g_{2} & \cdots & g_{K}\end{array}\right]^T.
\end{align}
The final estimation is then given by 
\begin{align}
\hat\phi & = f\left(\boldsymbol{\hat\phi}\right)  = \mathbf{g}^T \left(\mathbf{H}\boldsymbol{\hat\phi}+\mathbf{n}\right).\label{eq:summarizef}
\end{align}

\section{Calculating and Optimizing the Estimation Error\label{sec:QCstudy}}

The goal is for $f \left( \boldsymbol{\hat\phi} \right)$ to be an accurate estimator of $\phi$.  To measure the error in this estimation, we consider the mean squared error
\begin{align}
\epsilon \left( \boldsymbol{\hat\phi} \right) := &\; \mathbb{E} \left[\left( f\left(\boldsymbol{\hat\phi}\right)    - \phi\right)^2\right] \nonumber \\
 = &\;  \mathbb{E} \left[ f\left(\boldsymbol{\hat\phi}\right)^2\right] +  \phi^2   -     2\phi\mathbb{E}\left[ f\left(\boldsymbol{\hat\phi}\right)  \right] .
\end{align}
Of course, if $f$ is unbiased   then we have 
\begin{equation}
\epsilon \left( \boldsymbol{\hat\phi} \right)= \mathbb{V}\left[ f\left(\boldsymbol{\hat\phi}\right)\right]   ,
\end{equation}
however we do not consider this to be true in general.
The term $\mathbb{E} \left[ f\left(\boldsymbol{\hat\phi}\right) ^2 \right]$ is given by
\begin{align}
\mathbb{E} \left[ f\left(\boldsymbol{\hat\phi}\right) ^2 \right] &= \mathbf{g}^T \mathbf{H} \mathbf{R} \mathbf{H}^T \mathbf{g} +v \mathbf{g}^T \mathbf{g} , \label{eq:expoff}
\end{align}
where $\mathbf{R}$ represents the autocorrelation matrix of $\boldsymbol{\phi}$.
The term $\mathbb{E} \left[ f\left(\boldsymbol{\hat\phi} \right)  \right]$ is given by
\begin{align}
\mathbb{E} \left[ f\left(\boldsymbol{\hat\phi} \right) \right]  &=  \phi \mathbf{g}^T \mathbf{H} \mathbf{1}  .\label{eq:expf}
\end{align}
Combining these gives
\begin{align}
\epsilon \left( \boldsymbol{\hat\phi}  \right) = & \mathbf{g}^T \left(\mathbf{H}\mathbf{R} \mathbf{H}^T +\mathbb{I} v \right) \mathbf{g} + \phi^2  - 2 \phi^2\mathbf{1}^T \mathbf{H}^T \mathbf{g} .\label{eq:1234}
\end{align}
It is interesting to note that when $f$ is unbiased and $\phi\neq 0$, \eqref{eq:summarizef} and \eqref{eq:expf} give the constraint
\begin{align}
 \mathbf{g}^T \mathbf{H} \mathbf{1}  = 1.\label{eq:expfconstraint}
\end{align}

\subsection{Optimization}

In some scenarios, it may be possible to manipulate both $\mathbf{g}$ and $\mathbf{H}$ to minimize the error associated with our estimation. In other scenarios, the channel matrix $\mathbf{H}$ may be fixed\footnote{In this work, we assume that the function $\mathbf{g}$ can always be manipulated by the system designers.}. In the following lemma, we will establish the optimal values for $\mathbf{g}$ and $\mathbf{H}$ in the general setting. 
\begin{lem}\label{lem1}
The  optimal $\mathbf{g}$ (as a function of $\mathbf{H}$) that minimizes \eqref{eq:1234} is given by
\begin{align}
\mathbf{{g}}_\star \left( \mathbf{H} \right) = \phi^2  \left( \mathbf{H}\mathbf{R} \mathbf{H}^T + v\mathbb{I} \right)^{+} \mathbf{1} , \label{eq:gopt123}
\end{align}
while the optimal $\mathbf{H}$ (as a function of $\mathbf{g}$) is given by
\begin{align}
\mathbf{{H}_\star} \left( \mathbf{g} \right) =    \frac{\phi^2 }{ \mathbf{g}^T \mathbf{g} } \mathbf{g}  \mathbf{1}^T \mathbf{R} ^{-1},\label{eq:optimumH}
\end{align}
where $ \mathbf{A} ^{+}$ represents the pseudo inverse \cite{ben2003generalized} of $ \mathbf{A} $ for some square matrix $\mathbf{A}$.
\end{lem}
\begin{IEEEproof}
See Appendix \ref{applem}.
\end{IEEEproof}

As will be shown in the following (Lemma \ref{lemmaHopt2}), it is not possible for us to obtain a global optimum pair of solutions for $\mathbf{g}$ and $\mathbf{H}$ when noise is present. However, when noise is absent the global optimal solution can be achieved by arbitrarily fixing $\mathbf{H}$ and optimizing over $\mathbf{g}$ (or vice-versa). This is because when $v=0$, $\mathbf{g}$ and $\mathbf{H}$ always come as a pair in \eqref{eq:1234}. When $ \mathbf{1}$ is an eigenvector\footnote{Interestingly, it is easy to see this is a property of $\mathbf{R}$ when the elements of $\boldsymbol{\hat\phi}$ are i.i.d. } of $\mathbf{R}$ with eigenvalue 
\begin{equation}
\lambda = \frac{1}{N_e^2r}+K\phi^2,\label{eq:klambda}
\end{equation} 
an expression can be obtained for the minimum achievable error (this will be shown in Corollary~\ref{lem2}). Before presenting these ideas, we provide the following important remark.
\begin{rem}\label{remark1}
In general, the   optimal pair $\mathbf{g}_\star$ and $\mathbf{H}_\star$ provide a biased estimator since (from \eqref{eq:expfconstraint}, \eqref{eq:gopt123} and \eqref{eq:optimumH})
\begin{align}
\mathbf{{g}}_\star^T\mathbf{{H}_\star}\mathbf{1} = \phi^2 \mathbf{1}^T \mathbf{R}^{-1}\mathbf{1}.
\end{align}
However, when $ \mathbf{1}$ is an eigenvector of $\mathbf{R}$ with eigenvalue \eqref{eq:klambda} we have
\begin{align}
\mathbf{{g}}_\star^T\mathbf{{H}_\star}\mathbf{1} = \phi^2\lambda^{-1}K.
\end{align}
Consequently, in this case as $K$ grows large
\begin{align}
\mathbf{{g}}_\star^T\mathbf{{H}_\star}\mathbf{1} \to 1,
\end{align}
so that the estimator becomes unbiased in the limit.
\end{rem}

To understand why a global pair of solutions cannot be established in the noisy scenario, we must consider a particular consequence of Lemma \ref{lem1}. Specifically, from \eqref{eq:optimumH} we can see that 
\begin{align}
\mathbf{g}^T\mathbf{{H}_\star} \left( \mathbf{g} \right) =    \phi^2 \mathbf{1}^T \mathbf{R} ^{-1}.
\end{align}
Substituting this into \eqref{eq:1234}, we find that
\begin{align}
\min_{\mathbf{H}}\epsilon \left( \boldsymbol{\hat\phi}  \right) = &   \phi^4 \mathbf{1}^T \mathbf{R} ^{-1} \mathbf{1} + \mathbf{g}^T\mathbf{g}v + \phi^2 - 2\phi^4\mathbf{1}^T \mathbf{R} ^{-1}\mathbf{1}. \label{eq:minnedH}
\end{align}
 This then gives us the following lemma.
\begin{lem}\label{lemmaHopt2}
When the elements of $\boldsymbol{\hat\phi}$ are i.i.d. and $\mathbf{g} \neq \mathbf{0}$, we have 
\begin{align}
\min_{\mathbf{H}}\epsilon \left( \boldsymbol{\hat\phi}  \right) = &  v\mathbf{g}^T\mathbf{g} + \phi^2\left( \frac{\frac{1}{N_e^2 r}}{\frac{1}{N_e^2 r} + K\phi^2}\right).
\end{align}
Thus, the globaly optimal pair of solutions is achieved as $\mathbf{g}\to \mathbf{0}$.
\end{lem}
\begin{IEEEproof}
The result follows from \eqref{eq:minnedH} by noticing that $\mathbf{1}$ is an eigenvector of $\mathbf{R}$ with eigenvalue \eqref{eq:klambda}.
\end{IEEEproof}

\begin{cor}\label{lem2}
When the elements of $\boldsymbol{\hat\phi}$ are i.i.d. and  $v=0$, the global minimum error is given by
\begin{align}
 \min_{\mathbf{g},\mathbf{H}}\epsilon \left( \boldsymbol{\hat\phi}  \right) =  &\;  \phi^2 \left( \frac{ \frac{1}{N_e^2 r}}{\frac{1}{N_e^2 r} + K\phi^2}     \right) = \phi^2 \left( \frac{ 1 }{1 + {K\phi^2 N_e^2 r} }     \right)   .\label{eq:min28}
\end{align}
\end{cor}

At a high level, Lemma \ref{lemmaHopt2} is a somewhat intuitive result: if we are free to configure $\mathbf{H}$ arbitrarily, the best thing for us to do   is take $\mathbf{g}$ arbitrarily close to $\mathbf{0}$ (because this will suppress the noise). From \eqref{eq:gopt123}, this has the effect of `amplifying' the optimal $\mathbf{H}$ so that the wanted signal $\boldsymbol{\hat\phi}$ can propagate through the channel. Of course, if $\mathbf{g} = \mathbf{0}$ then neither the noise nor $\boldsymbol{\hat\phi}$ will be able to propagate through the channel. 

In practical scenarios, engineers may not be able to manipulate the channel matrix $\mathbf{H}$, instead only having access to $\mathbf{g}.$ In the following lemma, we establish the performance of the estimator when $\mathbf{H} = \mathbb{I}$. In this case, we have.
\begin{lem}\label{thm3}When the elements of $\boldsymbol{\hat\phi}$ are i.i.d. and $\mathbf{H} = \mathbb{I}$, the minimum error in the estimator is given by
\begin{align}
 \min_{\mathbf{g}}\epsilon \left( \boldsymbol{\hat\phi}  \right) =  &\;  \phi^2 \left( \frac{ \frac{1}{N_e^2 r}+v }{\frac{1}{N_e^2 r} + K\phi^2+v}     \right) .\label{eq:min1}
\end{align}
\end{lem}
\begin{IEEEproof}
See Appendix \ref{appTheorem2}.
\end{IEEEproof}

An interesting remark can now be made from the previous lemma and Lemma \ref{lemmaHopt2}.
\begin{rem}\label{remark2}
When minimizing over $\mathbf{H}$, we force $\mathbf{H}$ to have unit rank (see \eqref{eq:optimumH}). When minimizing over $\mathbf{g}$ we let $\mathbf{H}$ have full rank (see Lemma \ref{thm3}). In both scenarios, the error is given by
\begin{align}
\min_{\mathbf{H}\;\mathrm{or}\;\mathbf{g}}\epsilon \left( \boldsymbol{\hat\phi}  \right)   &=  
\Omega\left( \frac{1}{K N_e^2 r} \right).
\end{align}
when $v=0$.  Consequently, we must conclude that the asymptotic optimal performance is not uniquely determined by the rank of $\mathbf{H}$.
%When $v\neq0$, we see that
%\begin{align}
%\min_{ \mathbf{g}}\epsilon \left( \boldsymbol{\hat\phi}  \right)   &=  
%\Omega\left( \frac{1}{K  } \right),
%\end{align}
%while 
%\begin{align}
%\min_{ \mathbf{H}}\epsilon \left( 1 \right)   &=  
%\Omega\left( 1\right),
%\end{align}
\end{rem}

\section{A discussion\label{sec:study}}

The following points highlight important observations that can be made from the previous analysis.

First, for $f$ to be unbiased, when $\phi \neq 0$ \eqref{eq:expfconstraint} can be specialized in both scenarios. Specifically, we must have 
\begin{align}
 \mathbf{g}^T \mathbf{H}  \mathbf{1}  = 1. 
\end{align}
As was shown in Remark \ref{remark1}, $f$ becomes unbiased as $K\to\infty$. Also, when $\phi=0$ $f$ is unbiased.

Second, in practice, the minimum mean squared error (see \eqref{eq:min1}) will not be attainable. This is because the optimal parameterization of the system (i.e., how we configure $g$ and $c$) is dependent on the parameter that we are trying to estimate (i.e., $\phi$). This is clearly a non-causal scenario.

 Third, as was shown in Remark \ref{remark2}, the optimal performance of the system is not uniquely determined by the rank of $\mathbf{H}$. In particular, both unit rank and full rank $\mathbf{H}$ can achieve asymptotically optimal performance.

 Finally (and most critically), for the fixed $\mathbf{H}$ case,  because the optimal estimator can not be achieved in practice (see the second point of this discussion) the error will be given \emph{at best} by
\begin{align}
 \epsilon \left( \boldsymbol{\hat\phi}  \right)   &= \Omega\left( \frac{1}{K N_e^2 r} \right),
\end{align}
and this can only be achieved when $v = 0$. However, when  $v \neq 0$ \eqref{eq:min1} tell us that 
\begin{align}
 \epsilon \left( \boldsymbol{\hat\phi}  \right)   &= \Omega\left(  \frac{1}{K} \right).\label{eq:stiking}
\end{align}
This is a striking observation. It allows us to conclude that the $1/N_e^2r$ gain obtained from the individual quantum clusters will be lost \emph{entirely} by the classical processing if noise is present. Importantly, if the same number of resources ($N_e r K$ quantum states) had been employed, but combined over $N_e r K$ classical channels, the error would have decayed like \begin{equation}O\left(\frac{1}{N_e r K}\right).\end{equation} 
This observation highlights the important measures that must be put in place to ensure that entanglement assisted sensing provides its promised benefits.

\section{Conclusion\label{sec:conc}}

In this work, we presented and studied the quantum enhanced classical sensor network. We provided a general analysis of the system's optimal performance. We then made the critical observation that noise present within the estimator can severely degrade its performance. We also showed that in the limit as the number of classical channels grows large, the optimal estimator becomes unbiased. Future work will be performed to determine the effects of combining through quantum channels, rather than classical channels.

\section*{Acknowledgments}

This research was funded by the UK EPSRC (EP/K04057X/2) and the National Quantum Technologies Programme (EP/M01326X/1, EP/M013243/1).

\appendices

\section{Important Calculations}

The following calculations are used at various points in this work.
With  $\psi$ given by
\begin{equation}
\left| \psi \right> = \frac{1}{\sqrt{2}}\left( \left| 0 \right> + e^{\mathbf{i}\phi}\left| 1 \right>  \right)
\end{equation}
and 
\begin{equation}
X = \left| 0 \right>\!\left< 1 \right| + \left| 1 \right>\!\left< 0 \right| ,
\end{equation}
we have
\begin{align}
&\left< \phi \right| X \left| \phi \right> \\
=& \frac{1}{2}\left( \left< 0 \right| + e^{-\mathbf{i}\phi}\left< 1 \right|  \right)      \left(   \left| 0 \right>\!\left< 1 \right| + \left| 1 \right>\!\left< 0 \right|      \right)    \left( \left| 0 \right> + e^{\mathbf{i}\phi}\left| 1 \right>  \right)\nonumber\\
= &\cos \phi,\label{eq:basicExpectation}
\end{align}
and because $X^2 = \mathbb{I}$ we have
\begin{align}
\left< \phi \right| X^2 \left| \phi \right> =& 1.\label{eq:sigVar}
\end{align}
With $$\mathcal{X} = \left|\mathbf{0}_N\right>\!\left<\mathbf{1}_N\right| +  \left|\mathbf{1}_N\right>\!\left<\mathbf{0}_N\right|$$ and 
\begin{align}
\left| \psi_e\right>  = \frac{1}{\sqrt{2}} \left(  \left|\mathbf{0}_N\right>+ e^{\mathbf{i}N\phi} \left|\mathbf{1}_N\right> \right),
\end{align} 
we have
\begin{align}
\left< \psi _e\right|&\mathcal{X}  \left| \psi _e \right>\nonumber\\
 =& \frac{1}{2}\left( \left< \mathbf 0 \right| + e^{-N\mathbf{i}\phi}\left<\mathbf  1 \right|  \right)      \left(   \left|\mathbf  0 \right>\!\left<\mathbf    1 \right| + \left| \mathbf 1 \right>\!\left<\mathbf  0 \right|      \right)    \left( \left| \mathbf 0 \right> + e^{N\mathbf{i}\phi}\left| \mathbf 1 \right>  \right)\nonumber\\
= &\cos (N \phi),\label{eq:basicExpectation2}
\end{align}
and 
\begin{equation}
\left< \psi _e\right|\mathcal{X} ^2\left|\psi _e\right> = 1.\label{eq:SigVar}
\end{equation}

\section{Proof of Lemma \ref{lem1}\label{applem}}

The derivative of $\epsilon \left( \boldsymbol{\hat\phi}  \right)$ with respect to $\mathbf{g}$ is given by~\cite{petersen2008matrix}
\begin{align}
\frac{\partial\epsilon \left( \boldsymbol{\hat\phi}  \right)}{\partial\mathbf{g}} =   &  \;   2\mathbf{g}^T  \mathbf{H}\mathbf{R}  \mathbf{H}^T +  2v\mathbf{g}^T - 2 \phi^2   \mathbf{1}^T \mathbf{H}^T ,\label{eq:dedg}
\end{align}
while the derivative 
 with respect to $\mathbf{H}$ is given by  \cite{petersen2008matrix}
\begin{align}
\frac{\partial\epsilon \left( \boldsymbol{\hat\phi}  \right)}{\partial\mathbf{H}} =  &   \mathbf{R} \mathbf{H}^T  \mathbf{g} \mathbf{g}^T  + \mathbf{R} ^T\mathbf{H}^T  \mathbf{g} \mathbf{g}^T  \nonumber\\
&   - \mathbf{g} \mathbf{1}^T\phi^2   -   \mathbf{1}\phi^2 \mathbf{g}^T  .
\end{align}
From \eqref{eq:dedg},   the optimal $\mathbf{g}$ (as a function of $\mathbf{H}$) is given by
\begin{align}
\mathbf{{g}}_\star \left( \mathbf{H} \right) = \phi^2  \left( \mathbf{H}\mathbf{R} \mathbf{H}^T + v\mathbb{I} \right)^{+} \mathbf{1} ,  
\end{align}
while the optimal $\mathbf{H}$ (as a function of $\mathbf{g}$) is given by
\begin{align}
\mathbf{{H}_\star} \left( \mathbf{g} \right) =  \phi^2    \left( \mathbf{g} \mathbf{g}^T \right)^{+}   \mathbf{g}  \mathbf{1}^T \mathbf{R} ^{-1},
\end{align}
where $ \mathbf{A} ^{+}$ represents the pseudo inverse \cite{ben2003generalized} of $ \mathbf{A} $ for some square matrix $\mathbf{A}$. 

We can simplify $ \mathbf{H}_\star$ by considering the eigen/singular-value decomposition of $  \mathbf{g} \mathbf{g}^T $
\begin{equation}
\mathbf{g} \mathbf{g}^T = U\Lambda U^T.\label{eq:gg^T}
\end{equation}
We then have 
\begin{equation}
 \left(\mathbf{g} \mathbf{g}^T\right)^+  = U\Lambda^+U^T.\label{eq:gg^T+}
\end{equation}
 Since $\mathbf{g}$ is the only eigenvector of \eqref{eq:gg^T}  with eigenvalue $\left(\mathbf{g}^T\mathbf{g}\right)$, $\mathbf{g}$ is also the only eigenvector of  \eqref{eq:gg^T+}. The corresponding eigenvalue is $1/\mathbf{g}^T\mathbf{g}$. With this observation, $\mathbf{H}_\star$ becomes 
\begin{align}
\mathbf{{H}_\star} =  \frac{\phi^2\mathbf{g}  \mathbf{1}^T\mathbf{R} ^{-1}}{ \mathbf{g}^T\mathbf{g} }   .
\end{align}

\section{Proof of Theorem \ref{thm3}\label{appTheorem2}}

For this problem, from \eqref{eq:gopt123} and \eqref{eq:1234} we have
\begin{align}
& \min_{\mathbf{g}}\epsilon \left( \boldsymbol{\hat\phi}  \right) \nonumber \\
= &\; \phi^2 - \phi^2 \mathbf{1}^T \mathbf{{g}_\star} \nonumber\\
= &\; \phi^2 \mathbf{1}^T \left( \mathbb{I}/K -  \left( \mathbf{R} + v\mathbb{I} \right)^{-1} \phi^2 \right)  \mathbf{1} \nonumber \\
= &\;  \frac{ \phi^2 }{K} \mathbf{1}^T \left( 	  \mathbf{R}  + v\mathbb{I} \right)^{-1} \left(    \mathbf{R} + v\mathbb{I}   - K\phi^2\mathbb{I} \right)  \mathbf{1}   ,\label{eq:min134}
\end{align}
where the first and second equalities follow by substituting \eqref{eq:gopt123} into the equation.

As with before, equation \eqref{eq:min1} can be dealt with by noticing that, if the elements of $\boldsymbol{\hat\phi}$ are i.i.d., $\mathbf{1}$ becomes an eigenvector of $ \mathbf{R} $ and $ \mathbf{R} ^{-1}$ with eigenvalues $\lambda$ and $1/\lambda$ (see \eqref{eq:klambda}).
From \eqref{eq:min134}, the result then follows.

\bibliographystyle{plain}
\bibliography{QCsensor}

\end{document}